\documentclass[letter,twocolumn]{jpsj3}

\usepackage{bm}
\usepackage{url}

\newcommand{\ket}[1]{|#1\rangle}
\newcommand{\bra}[1]{\langle #1|}

\title{Attention-Like Hebbian Learning from Quantum Probability Flow and Quantum-Annealer Tests}

\author{Masayuki \textsc{Ohzeki}$^{1,2,3,4}$\thanks{Corresponding author. mohzeki@tohoku.ac.jp}}

\inst{$^{1}$Graduate School of Information Sciences, Tohoku University, Sendai 980-8579, Japan \\
$^{2}$Department of Physics, Institute of Science Tokyo, Tokyo 152-8550, Japan \\
$^{3}$Research and Education Institute for Semiconductors and Informatics, Kumamoto University, Kumamoto 860-8555, Japan \\
$^{4}$Sigma-i Co., Ltd., Tokyo 108-0075, Japan}

\abst{We propose a quantum probability-flow principle for deriving local learning rules in associative memory. A transverse field defines leakage channels from data states, and minimizing the measured survival loss gives stability-driven updates. For imaginary-time, dephased dynamics, the local leakage free energy is the log-sum-exp of energy gaps; its gradient is a softmax-weighted Hebbian rule. Real-time stability instead yields a power-law weighting. D-Wave standard- and fast-anneal tests of a one-hot attention forward map are better fitted by an effective softmax than by a Lorentzian power law.}

\kword{quantum probability flow, Hebbian learning, attention mechanism, quantum annealing, Ising model}

\begin{document}
\maketitle

Hebbian learning and attention mechanisms represent two influential ideas for storing and retrieving information in neural systems. Hebbian learning reinforces local correlations and underlies the classical Hopfield memory,\cite{Hebb1949,Hopfield1982} whereas the softmax attention used in Transformers performs context-dependent weighting of candidate information.\cite{Vaswani2017} Their relation has become sharper through modern Hopfield networks and dense associative memories, where attention can be interpreted as retrieval from an energy landscape.\cite{Krotov2016,Ramsauer2020} Despite this proximity, the two principles are usually motivated in different languages: Hebbian learning as a local synaptic rule, and attention as differentiable selection over scores. A compact physical derivation of the interpolation between them is therefore useful.

In this Letter, we show that a common local stability principle produces both the Hebbian limit and an attention-like finite-temperature correction. The principle is a quantum extension of minimum probability flow (MPF),\cite{SohlDickstein2011} in which a weak transverse field specifies the directions along which probability can leak out of a stored pattern. MPF is not an alternative model to a Boltzmann machine; it is a partition-function-free local learning principle for the same energy-based distribution. In classical MPF, one chooses local stochastic dynamics whose equilibrium distribution is the model distribution, initializes the system at the empirical data distribution, and minimizes the initial probability flow away from the data. This replaces global likelihood learning by a local escape-rate criterion while preserving the Boltzmann-machine energy landscape. The quantum version below keeps this local philosophy but lets the driver Hamiltonian define the error modes.

The central question is not how to reproduce the full architecture of a Transformer, but how softmax weighting can arise from a microscopic stability criterion. The answer is simple: imaginary-time probability flow gives a local log-sum-exp free energy over possible errors, and the gradient of this free energy is a softmax-weighted Hebbian update. This gives a statistical-physics interpretation of attention as selective repair of unstable memory directions.

Consider an Ising memory with
\begin{equation}
H
=-\sum_{i<j}J_{ij}\sigma_i^z\sigma_j^z-\sum_i h_i\sigma_i^z ,
\label{eq:ham}
\end{equation}
where the couplings $J_{ij}$ and local fields $h_i$ are the trainable energy parameters. A data pattern is a computational basis state $\ket{\bm{x}}$ with $x_i=\pm1$. The $\mu$th data pattern is denoted by $\bm{x}_{\mu}$, with components $x_{\mu i}$, and the data set is represented by
\begin{equation}
\rho_{\rm data}=\frac{1}{M}\sum_{\mu=1}^{M}
\ket{\bm{x}_{\mu}}\bra{\bm{x}_{\mu}} .
\end{equation}
We introduce the driver $V=\Gamma\sum_i\sigma_i^x$, which connects a pattern $\bm{x}$ to its single-spin-flipped neighbors $\bm{x}^{(k)}$. The associated local energy gap is
\begin{equation}
\Delta E_k(\bm{x})
=E(\bm{x}^{(k)})-E(\bm{x})
=2x_k\left(h_k+\sum_{l(\ne k)}J_{kl}x_l\right).
\label{eq:gap}
\end{equation}
For memory storage, positive gaps mean that every single flip raises the energy of the stored pattern. Thus $\Delta E_k$ is a local margin or stability gap. Ordinary Hebbian learning increases average correlations in the data, but it does not distinguish a spin sitting behind a large barrier from a spin about to flip. The probability-flow view below makes this distinction explicit.

The quantum probability-flow loss is defined from the probability of recovering the original pattern after evolution by $H+V$ and projective measurement in the computational basis. This measurement step turns the quantum evolution into a stability test for stored classical memories: after a short perturbation, does a measurement return the original pattern, or one of its nearby errors? In the imaginary-time, dephased, single-flip approximation,\cite{Suzuki1976} the transverse field fixes the allowed leakage channels and the imaginary-time dynamics assigns a detailed-balance weight
\begin{equation}
K_{\beta}(\bm{x}^{(k)}|\bm{x})
=\Gamma^2\exp[-\beta \Delta E_k(\bm{x})],
\label{eq:flow}
\end{equation}
where $\beta$ is the effective inverse temperature. Equation (\ref{eq:flow}) may be regarded as the local quantum-MPF transition weight: it keeps the neighborhood structure generated by the transverse field, while the imaginary-time filter exponentially suppresses transitions that climb a high energy barrier. The normalized survival probability is then
\begin{equation}
{\cal F}_{\beta}(\bm{x})
=\frac{1}{1+\sum_k K_{\beta}(\bm{x}^{(k)}|\bm{x})},
\end{equation}
and the local probability-flow loss becomes
\begin{equation}
\ell_{\beta}(\bm{x})
=-\log{\cal F}_{\beta}(\bm{x})
=\log\left[1+\Gamma^2\sum_k e^{-\beta\Delta E_k(\bm{x})}\right].
\label{eq:lse}
\end{equation}
Thus the loss is a local free energy of the leakage channels. The log-sum-exp structure is the key point: its gradient automatically generates a softmax distribution over the most fragile directions around the pattern.

The variational meaning of this step is also useful. Ignoring the no-flip constant in Eq. (\ref{eq:lse}), define the soft minimum of the local gaps by
\begin{equation*}
\begin{split}
\Phi_{\beta}(\bm{x})
&=-\frac{1}{\beta}\log\sum_k e^{-\beta\Delta E_k(\bm{x})} \\
&=\min_{q\in\Delta_N}\Biggl[
\sum_k q_k\Delta E_k(\bm{x})
+\frac{1}{\beta}\sum_k q_k\log q_k
\Biggr],
\end{split}
\end{equation*}
where $\Delta_N$ is the probability simplex over leakage channels. The minimizing distribution is
\begin{equation}
q_k(\bm{x})
=\frac{e^{-\beta\Delta E_k(\bm{x})}}
{\sum_l e^{-\beta\Delta E_l(\bm{x})}} .
\label{eq:q}
\end{equation}
Therefore the softmax is not only a Boltzmann factor; it is the optimal entropy-regularized distribution of probability flow over possible errors. High temperature spreads the flow almost uniformly, whereas low temperature concentrates it on the weakest gap.

This observation identifies the physical origin of the attention-like structure. The driver Hamiltonian first creates competing candidates, here the single-spin-flipped states $\{\bm{x}^{(k)}\}$. The energy gaps assign scores to these candidates, with $-\Delta E_k$ measuring how dangerous the $k$th leakage channel is. Imaginary-time evolution converts the scores into exponential weights, and the normalization of the measured survival probability converts these weights into a softmax distribution. Finally, taking the gradient of the local free energy writes the stabilizing response into the parameters. In short, the chain of operations is
$V \Rightarrow \{\hbox{channels}\}$, $\Delta E_k \Rightarrow \{\hbox{scores}\}$, and $q_k\propto e^{-\beta \Delta E_k} \Rightarrow \{\hbox{softmax weights}\}$ followed by the parameter update that plays the role of value aggregation. 
Thus the attention-like ingredient is not imposed as an architecture; it appears from normalized competition among quantum leakage channels.

Indeed, defining
\begin{equation}
r_k(\bm{x})
=\frac{\Gamma^2 e^{-\beta\Delta E_k(\bm{x})}}
{1+\Gamma^2\sum_l e^{-\beta\Delta E_l(\bm{x})}},
\label{eq:r}
\end{equation}
we obtain
\begin{align}
-\frac{\partial \ell_{\beta}}{\partial J_{ij}}
&=2\beta [r_i(\bm{x})+r_j(\bm{x})]x_ix_j,\\
-\frac{\partial \ell_{\beta}}{\partial h_i}
&=2\beta r_i(\bm{x})x_i .
\end{align}
Consequently, gradient descent on the data-averaged loss
$M^{-1}\sum_\mu \ell_{\beta}(\bm{x}_{\mu})$ yields
\begin{equation}
\delta J_{ij}\propto
\frac{1}{M}\sum_{\mu=1}^{M}
[r_i(\bm{x}_{\mu})+r_j(\bm{x}_{\mu})]x_{\mu i}x_{\mu j} .
\label{eq:update}
\end{equation}
This is a Hebbian outer-product update whose strength is modulated by the local leakage weights. The factor $r_i+r_j$ means that the correlation $x_ix_j$ is reinforced mainly when one of its endpoint spins is locally unstable. This is exactly the behavior expected from a memory-repair rule: learning effort is not spent uniformly on all stored correlations, but concentrated near weak barriers in the energy landscape. Once the instability weights have been computed from the local fields, the synaptic increment depends only on the two spin values and endpoint fragilities.

Equation (\ref{eq:q}) identifies $q_k$ as a softmax attention distribution over the spin-flip channels. The pattern $\bm{x}$ plays the role of the query, the possible flips are the keys, the score is $-\Delta E_k$, and the value written into the memory is the correlation $x_ix_j$. This is not an architectural identity with Transformer self-attention; rather, it is a local physical mechanism that produces the same softmax weighting principle. The present model has no learned query, key, and value projection matrices. Instead, it derives attention weights from the stability margins of an Ising memory. The resulting update is therefore best described as an attention-like Hebbian rule.

The analogy can be made more explicit at the level of gradients. A standard attention layer forms a weighted sum $\sum_k {\rm softmax}(s)_k V_k$, where ${\rm softmax}(s)_k=e^{s_k}/\sum_l e^{s_l}$, from scores $s_k$ and values $V_k$. Here the score is fixed by physics, $s_k=-\beta\Delta E_k$, and the value is the stabilizing direction $\partial\Delta E_k/\partial\eta$, where $\eta$ denotes one of the energy parameters $J_{ij}$ or $h_i$. Therefore the learning signal has the same mathematical form,
\begin{equation}
-\partial_{\eta}\ell_{\beta}
=\beta\sum_k r_k(\bm{x})
\partial_{\eta}\Delta E_k(\bm{x}),
\label{eq:attention-gradient}
\end{equation}
apart from the no-flip normalization contained in $r_k$. This is why the softmax is not a superficial resemblance. It is the derivative of the leakage free energy, and the value vectors are the parameter changes that suppress each error mode.

This also explains why the Hebbian factor appears at the same time. The probability flow selects which error channel should be repaired, but the Ising parametrization determines how that repair is written into the model. Since $\partial\Delta E_k/\partial J_{ij}$ is proportional to the local product $x_ix_j$ when the flipped site touches the edge $(i,j)$, the gradient of the flow loss necessarily writes correlations. Thus the causal chain is: the KL/MPF objective defines an information leakage to be minimized, the Gibbs variational principle turns leakage competition into softmax attention, and the pairwise energy model turns the stabilizing gradient into a Hebbian correlation update.

The limiting cases are immediate. In the high-temperature limit $\beta\to0$, $q_k\to1/N$ and the pattern-dependent prefactor $a(\bm{x})$ becomes a constant. Equation (\ref{eq:update}) therefore reduces, up to an overall learning rate, to the Hebbian rule
\begin{equation}
\delta J_{ij}\propto
\sum_{\mu=1}^{M}x_{\mu i}x_{\mu j} .
\end{equation}
Thus ordinary Hebbian learning corresponds to isotropic reinforcement, where thermal fluctuations do not distinguish strong and weak local barriers. In the low-temperature limit, $q_k$ concentrates on the minimum gap. The update then repairs only the most unstable spin direction, giving a hard-attention limit of the same stability principle.

This interpolation also clarifies why the temperature parameter is not a technical decoration. At high temperature, the local environment is effectively unresolved, so all directions are treated equally and the rule becomes a one-shot correlation memory. At finite temperature, learning becomes margin-aware: the stored pattern itself determines which neighboring errors should be suppressed most strongly. At very low temperature, only the weakest margin matters, and the rule becomes greedy repair of the closest escape channel. In this sense, attention is not added to Hebbian learning by hand; it appears when the stability loss is made sensitive to the spectrum of local errors.

The same formulation also suggests a practical diagnostic. If a set of patterns has already been stored by a Hebbian rule, the gaps $\Delta E_k(\bm{x}_{\mu})$ reveal which bits and which memories are poorly protected. Applying Eq. (\ref{eq:update}) selectively increases the fragile gaps while leaving already stable directions almost unchanged. Thus the proposed rule can be interpreted either as a learning principle or as a post-Hebbian refinement procedure for reshaping the local margins of an associative memory.

Real-time evolution leads to a different weighting. For a two-level channel $\ket{\bm{x}}\leftrightarrow\ket{\bm{x}^{(k)}}$, perturbation theory gives a Rabi transition probability bounded by
\begin{equation}
P_{\bm{x}\to\bm{x}^{(k)}}(t)
\leq \frac{4\Gamma^2}{[\Delta E_k(\bm{x})]^2}
\end{equation}
when $\Gamma\ll|\Delta E_k|$. A time-independent robust objective is therefore proportional to
$\sum_k[\Delta E_k(\bm{x})^2+\varepsilon^2]^{-1}$, with $\varepsilon$ a small regularizer. For positive gaps this gives
\begin{equation}
\delta J_{ij}\propto
\left[
\frac{\Delta E_i}{(\Delta E_i^2+\varepsilon^2)^2}
+
\frac{\Delta E_j}{(\Delta E_j^2+\varepsilon^2)^2}
\right]x_ix_j ,
\end{equation}
which approaches a power-law weight $\sim(\Delta E)^{-3}$ as $\varepsilon\to0$. Coherent quantum stability therefore emphasizes small gaps algebraically rather than exponentially, and is more singular than the imaginary-time softmax rule.

The contrast between the imaginary-time and real-time results is physically instructive. Imaginary-time evolution acts as a dissipative or thermal filter and produces normalized competition among leakage channels. Real-time evolution preserves phase coherence and is controlled by resonance denominators. If the measured transition probabilities themselves are normalized and used as an input-output map, the weights are oscillatory in time and, after dephasing or time averaging, become Lorentzian or power-law weights rather than exponential softmax weights. Thus a coherent real-time device does not directly implement standard Transformer attention, but a response governed by resonant leakage. The softmax rule is the thermal/free-energy counterpart, while the power-law rule is the coherent-stability counterpart. This distinction is consistent with the broader literature on non-softmax attention kernels, including kernelizable attention, Gaussian and $L_2$ attention, sparse entmax attention, and power-law-biased attention.\cite{Choromanski2021,Chen2021Skyformer,Kim2021Lipschitz,Correia2019,Hegazy2026}

These results show that Hebbian learning and attention-like weighting can be viewed as two regimes of a single local stability principle. A transverse-field driver probes single-spin robustness and produces node-wise attention over fragile directions; other drivers would probe other structures. For example, a two-spin driver $\sum_{ij}\sigma_i^x\sigma_j^x$ would assign weights to correlation-flip channels and naturally lead to an edge-attention rule, reminiscent of graph attention networks.\cite{Velickovic2018} More generally, the driver determines the error modes, and the probability-flow loss determines how learning resources are allocated among them.

This perspective also suggests a physical meaning of multiple heads. If the driver is decomposed as $V=\sum_{\alpha}\Gamma_{\alpha}V_{\alpha}$, each component probes a different class of leakage channels and generates its own local softmax distribution. A ``head'' is then a particular stability test: one head may inspect single-bit errors, another pair flips, and another structured collective moves. Multi-head attention can therefore be interpreted, at this abstract level, as the simultaneous monitoring of several noise or error geometries.

This interpretation suggests a route toward hardware-native attention. Present Transformer implementations on digital accelerators explicitly form scores, apply softmax normalization, and multiply by values. In the present view, the learning stage may be regarded as Boltzmann-machine or MPF-type training of $J_{ij}$ and $h_i$, possibly using a quantum annealer or a stochastic Ising machine as an approximate sampler.\cite{Kadowaki1998,Johnson2011,Marshall2019} After these parameters are fixed, a thermal or open-system Ising device can be used as a forward processor: candidate generation, Boltzmann weighting, and normalization are supplied by physical relaxation or sampling. A coherent real-time device would instead realize the power-law response discussed above. Thus the possible hardware advantage is not a direct replacement of dense GPU linear algebra, but a reformulation of forward computation as a physical free-energy or response-function evaluation process.

It is useful to separate two levels of this proposal. The theory above is intrinsically interacting: the couplings $J_{ij}$ define the memory landscape, and their gradients give the Hebbian part of the update. A QPU can therefore in principle implement forward maps in which candidate channels are not independent but interact through an energy function. Such interacting attention would go beyond the usual Transformer softmax over independent scores. The experiment below deliberately tests the simpler noninteracting limit first: whether QPU final readout can reproduce an ordinary softmax-like Transformer forward map before exploiting the more distinctive interacting capability of the hardware.

We performed a small hardware test of this noninteracting baseline on a D-Wave Advantage quantum annealer. For a query $\bm{q}$, keys $\bm{k}_m$, and scores $s_m=\bm{q}\cdot\bm{k}_m/\sqrt{d}$, we encoded a one-hot attention selector by
\begin{equation}
E_{\bm{q}}(\bm{z})
=A\left(\sum_m z_m-1\right)^2
-\lambda\sum_m s_m z_m ,
\label{eq:onehot}
\end{equation}
where $z_m\in\{0,1\}$. Conditioned on valid one-hot samples, an ideal thermal sampler gives
$p(m|\bm{q})\propto \exp(\beta_{\rm eff}\lambda s_m)$, i.e., a physical softmax forward map. We fitted the measured conditional distribution by $p_{\rm sm}(m|\bm{q};\kappa)\propto\exp(\kappa s_m)$, where $\kappa=\beta_{\rm eff}\lambda$ for an ideal thermal sampler. In runs with eight keys, eight queries, and 1000 reads per query, this slope tracked the programmed score scale: $\kappa/\lambda=2.10,2.17,2.11$ for $\lambda=1,2,3$, respectively. The global KL divergences to the softmax family were also smaller than those to the Lorentzian/power-law surrogate, supporting an effective free-energy attention rather than a coherent response kernel.

We also tested the fast-anneal protocol, which can access nanosecond anneal times and has been used to study coherent annealing dynamics.\cite{King2022,DWaveDocsSolverParameters} Since fast annealing requires zero direct Ising fields, the same energy was rewritten in spin variables and the linear terms were implemented using ancilla qubits with flux-bias offsets, following the D-Wave prescription.\cite{DWaveDocsAnnealing} The results are summarized in Table \ref{tab:dwave}. Fast annealing strongly reduced the fitted softmax slope compared with a 20 $\mu$s standard anneal, but in the global-KL test it did not make the final-readout distribution better described by the power-law model.

\begin{table}[t]
\caption{D-Wave forward tests for the spin formulation of Eq. (\ref{eq:onehot}) with four keys, eight queries, 1000 reads per query, $A=0.5$, and $\lambda=2.0$. KL$_{\rm sm}$ and KL$_{\rm pl}$ denote global KL divergences to the fitted softmax and power-law families; smaller values indicate better fits. The parameter $\kappa$ is the fitted softmax slope in $p_{\rm sm}(m|\bm{q};\kappa)\propto\exp(\kappa s_m)$.}
\label{tab:dwave}
\begin{center}
\scriptsize
\begin{tabular}{cccccc}
\hline
protocol & $t_a$ ($\mu$s) & valid & KL$_{\rm sm}$ & KL$_{\rm pl}$ & $\kappa$\\
\hline
fast & 0.005 & 0.770 & 0.00765 & 0.02179 & 5.58\\
fast & 0.007 & 0.806 & 0.01151 & 0.02407 & 6.55\\
fast & 0.05 & 0.917 & 0.02155 & 0.03917 & 10.28\\
fast & 0.5 & 0.842 & 0.01653 & 0.02981 & 9.67\\
standard & 20 & 0.943 & 0.00517 & 0.01508 & 30.85\\
\hline
\end{tabular}
\end{center}
\end{table}

These data should be regarded as proof-of-principle rather than final benchmarking: gauge averaging, chain-strength sweeps, readout-error analysis, and larger random instances remain necessary. Nevertheless, the qualitative message is clear. Shortening the anneal toward the fast regime changes the effective temperature and validity of the one-hot forward map, but in this final-readout test the global KL remains closer to a softmax/free-energy family than to the Lorentzian response expected from isolated coherent real-time transition probabilities. Thus D-Wave-like annealers appear naturally suited to hardware-native Boltzmann attention, whereas genuine response-function attention would require more direct access to coherent transition probabilities. The next natural step is assessing interacting QPU forward maps as new Transformer-like architectures.

\section*{Acknowledgment}
This study was supported by the Cross-ministerial Strategic Innovation Promotion Program (SIP) from the Cabinet Office (No. 23836436).

\end{document}